\documentclass{sf2a-conf2011}
\usepackage{graphicx}
\usepackage{hyperref}
\usepackage[]{natbib}  
\usepackage[cyr]{aeguill}
\usepackage{epstopdf}

\def\BibTeX{{\rm B\kern-.05em{\sc i\kern-.025em b}\kern-.08em
    T\kern-.1667em\lower.7ex\hbox{E}\kern-.125emX}}
\bibpunct{(}{)}{;}{a}{}{,}  

\begin{document}

\TitreGlobal{SF2A 2011}


\title{LBG properties from z$\sim$3 to z$\sim$6}

\runningtitle{LBG properties from z$\sim$3 to z$\sim$6}

\author{S. de Barros}\address{Observatoire de Gen\`eve, Universit\'e de Gen\`eve, 51, Ch. des Maillettes, CH-1290 Versoix, Switzerland}

\author{D. Schaerer$^{1,}$}\address{Laboratoire d'Astrophysique de Toulouse-Tarbes, Universit\'e de Toulouse, CNRS, 14 Avenue E. Belin, 31400 Toulouse, France}

\author{D. P. Stark}\address{Kavli Institute of Cosmology and Institute of Astronomy, University of Cambridge, Madingley Road, Cambridge CB30HA, UK}


\setcounter{page}{93}

\index{de Barros, S.}
\index{Schaerer, D.}
\index{Stark, D. P.}


\maketitle


\begin{abstract}
We analyse the spectral energy distribution (SED) of $U$, $B$, $V$ and $i$-dropout samples from GOODS-MUSIC and we determine their physical properties, such as stellar age and mass, dust attenuation and star formation rate (SFR). Furthermore, we examine how the strength of Ly$\alpha$ emission can be constrained from broad-band SED fits instead of relying in spectroscopy. We use our SED fitting tool including the effects of nebular emission and we explore  different star formation histories (SFHs). We find that SEDs are statistically better fitted with nebular emission and exponentially decreasing star formation. Considering this result, stellar mass and star formation rate (SFR) estimations modify the specific SFR (SFR/M$_{\star}$) - redshift relation, in comparison to previous studies. Finally, our inferred Ly$\alpha$ properties are in good agreement with the available spectroscopic observations.
\end{abstract}



\section{Introduction}
Understanding how galaxies assemble their mass is an important goal in high redshift studies \citep{starketal2009}. Using new SED modeling techniques taking into account nebular emission \citep{schaerer&debarros2009,schaerer&debarros2010}, we present preliminary results of a study of a large sample of high redshift galaxies. We find that statistically, the model which provides the best fit is a model including nebular emission, with an exponentially decreasing star formation history. This implies a new estimation on age, stellar mass and SFR. Also, we here demonstrate that the properties of Ly$\alpha$  can also be inferred from broad-band observations, at least statistically for large samples. This allows us, for example, to determine trends of Ly$\alpha$ with redshift and other parameters, without resort to spectroscopy.

We adopt a $\Lambda$-CDM cosmological model with $H_{0}$=70 km s$^{-1}$ Mpc$^{-1}$, $\Omega_{m}$=0.3 and $\Omega_{\Lambda}$=0.7. 
  
\section{Data and method}
We have used the GOODS-MUSIC catalog of \cite{santinietal2009} for the GOODS-South field, providing photometry in the $U$, $B_{435}$, $V_{606}$, $i_{776}$, $z_{850}$, $J$, $H$, $K$, bands mostly from the VLT and HST, and the 3.6, 4.5, 5.8 and 8.0 $\mu$m bands from the IRAC camera onboard $Spitzer$. Using standard criteria as in \cite{starketal2009}, we have selected $U$, $B$, $V$ and $i$-drop galaxies. To reduce the contamination rate, we have only retained the objects whose median photometric redshifts agree with the targetted redshift range. We are thus left with a sample of 389, 705, 199 and 60 galaxies at $z\sim3$, $z\sim4$, $z\sim5$ and $z\sim6$.

Our SED fitting tool as already being described in \cite{schaerer&debarros2009} and \cite{schaerer&debarros2010}, here, we briefly summarize.

We use a recent, modified version of the Hyperz photometric redshift code of \cite{bolzonellaetal2000}, taking into account nebular emission (lines and continua) which can impact broad brand photometry and derived properties \citep{schaerer&debarros2009,schaerer&debarros2010,onoetal2010}. We consider a large set of spectral templates \citep{bruzual&charlot2003}, covering different metallicities and a wide range of star formation (SF) histories (exponentially decreasing, constant and rising SF) and we add the effects of nebular emission.

We adopt a Salpeter IMF \citep{salpeter1955} from 0.1 to 100 M$_\odot$, and we 
properly treat the returned ISM mass from stars. IGM is treated following \cite{madau1995} and the extinction is described by the Calzetti law \citep{calzettietal2000}. Nebular emission from continuum processes and lines is added to the spectra
predicted from the GALAXEV models as described in 
\cite{schaerer&debarros2009}.
We define three models:

\begin{itemize}
	\item Reference model: constant star formation, age $>$ 50 Myr and solar metallicity.
	\item Decreasing model: exponentially decreasing star formation (SFR $\propto \exp(-t/\tau)$). Metallicity and $\tau$ are free parameters, $\tau$ varying from 10 Myr to $\infty$ (SFR = constant).
	\item Rising model: we use the mean rising star-formation history from
the simulations of \cite{finlatoretal2011}.  Metallicity is a free parameter.
\end{itemize}
Furthermore, we define two options: first, +NEB which indicates that we include nebular emission, with all lines except  Ly$\alpha$, since this line may be attenuated by radiation transfer processes inside the galaxy or by the intervening intergalactic medium and second, +NEB+Ly$\alpha$ which indicates that we include nebular emission, with all lines.
In all cases, we consider z $\epsilon$ [0,10] in steps of 0.1 and A$_V$ $\epsilon$ [0,4] mag in steps of 0.1.

To determine properly confidence intervals for all the parameters, we ran 1000 Monte Carlo simulations for each object by perturbing the input broadband photometry assuming the photometric uncertainties are Gaussian. This procedure provides the probability distribution of the physical parameters for each source and for the ensemble of sources.

\section{Fit quality}

\begin{figure}[ht!]
\centering
 \includegraphics[width=0.48\textwidth,trim=2cm 0cm 4cm 14cm,clip]{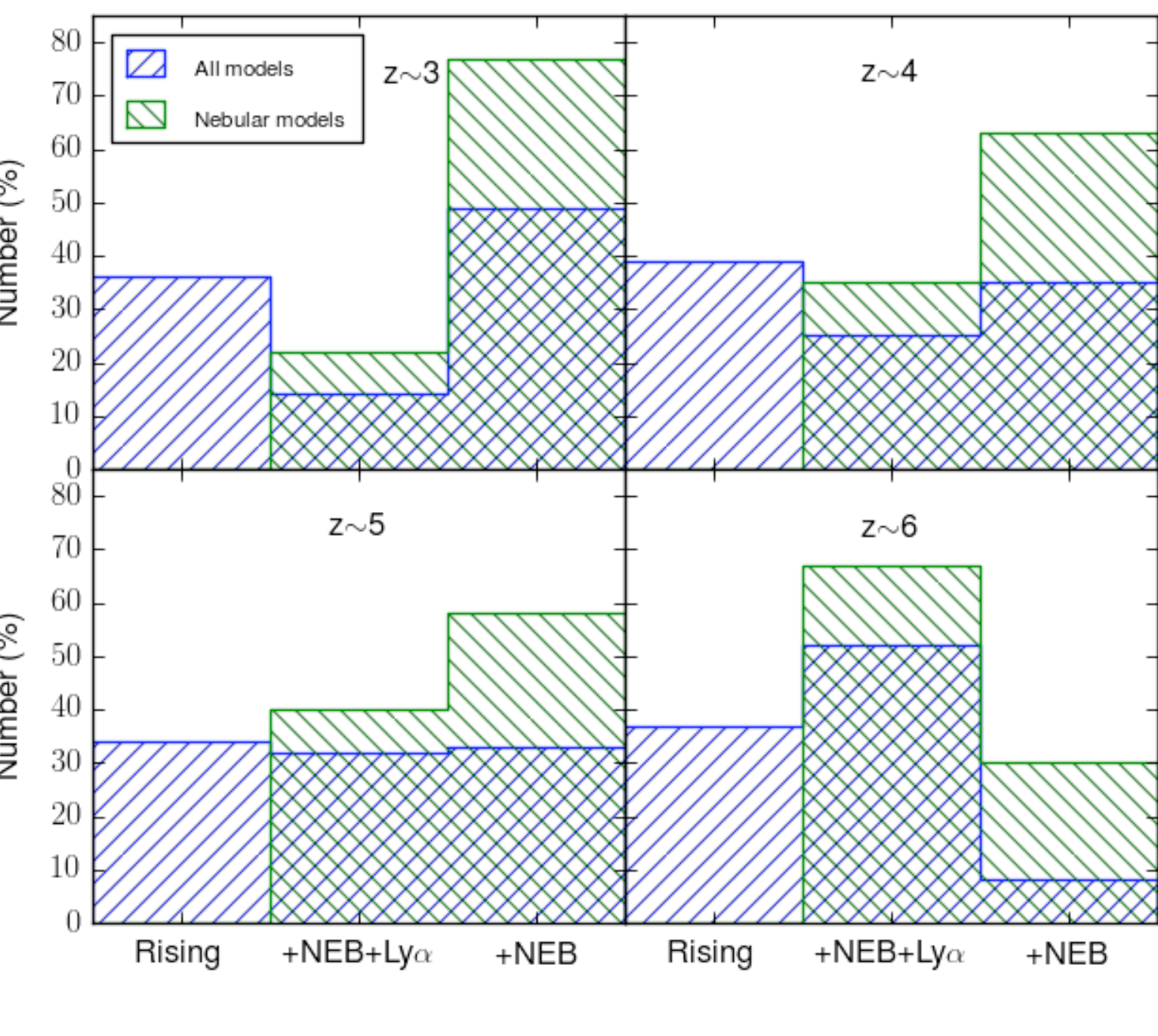}
 \includegraphics[width=0.48\textwidth,trim=1cm 0cm 3.25cm 14cm,clip]{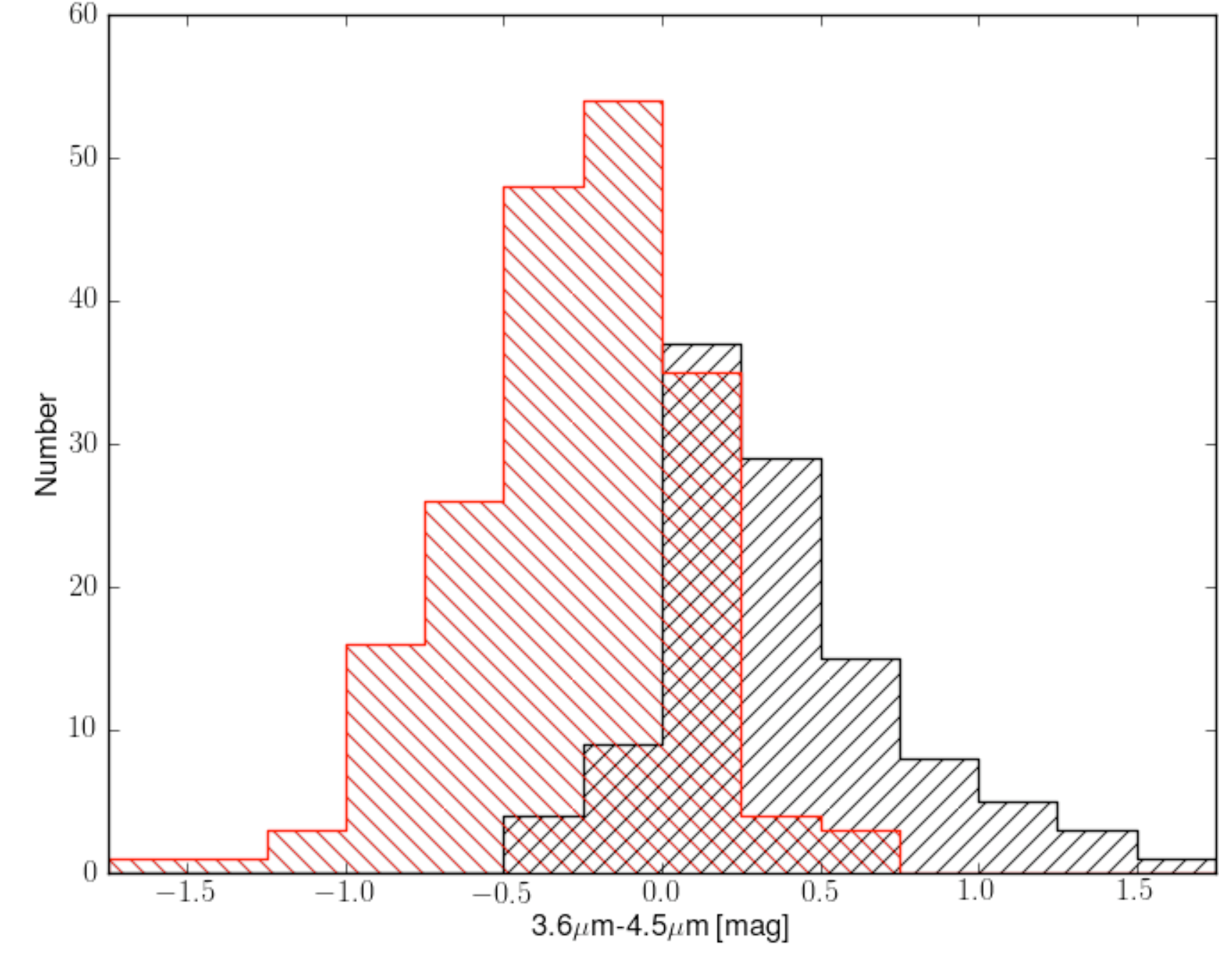} 
  \caption{{\bf Left:} In blue: best fits distribution among Rising, Rising+NEB+Ly$\alpha$ and Rising+NEB model and in green: best fits distribution between Rising+NEB+Ly$\alpha$ and Rising+NEB model. {\bf Right:} 3.6$\mu$m-4.5$\mu$m color histogram for a sub-sample of z  $\epsilon \left[3.8,5\right]$ objects. In red, best fitted objects with nebular emission
   and black, best fitted objects without nebular emission.}
  \label{debarros:fig1}
\end{figure}

For the four samples and each SFH (see figure \ref{debarros:fig1} for Rising model, we find similar results for all SFHs), $\sim$35\% (30\%-39\%) of the objects are best fitted without taking into account nebular emission. This fraction is independent of parameters like M$_{1500}$ or the filter's number available. For the three SFH, in $U$, $B$, $V$ and $i$-dropout, we found respectively 68\%, 71\%, 71\% and 77\% common objects best fitted without nebular emission and 85\%, 80\%, 94\% and 88\% common objects best fitted taking into account nebular emission.

H$\alpha$ is a strong line at 656.4 nm (reference frame) and must affect 3.5$\mu$m-4.6$\mu$m color for objects between $z$=3.8 and $z$=5 \citep{shimetal2011}. We selected
$B$-dropout objects with 3.6$\mu$m and 4.5$\mu$m avalaible data (excluding non-detections) and with median redshit between 3.8 and 5. We obtain a
subsample of 303 objects, with similar distribution for the best fits ($\sim$35\%-$\sim$65\%). Figure \ref{debarros:fig1} shows that  objects best fitted with nebular emission are almost all blue objects, which can be easily explain by strong nebular emission.
We define two categories of objects: Wneb, which are objects best fitted without nebular emission and Sneb, which are objects best fitted with nebular emission, choosing arbitrarily the "Decreasing model" to do this selection, considering the similarity between these two categories among the SFHs.

Wneb objects show a slight $\chi^2_r$ improvement (15\% to 25\% lower) for models without nebular emission in comparison with the $\chi^2_r$ for models including nebular emission. At the opposite, Sneb objects show a large improvement of the $\chi^2_r$ for models including nebular emission (36\% to 51\% lower). In all cases, we find lowest $\chi^2_r$ with Decreasing/Rising models and Sneb are significantly best fitted with Decreasing+NEB/+NEB+Ly$\alpha$ models.

\section{Star formation rate and stellar mass}

\begin{figure}[ht!]
\centering
 \includegraphics[width=0.7\textwidth,trim=1cm 8cm 2cm 8.5cm,clip]{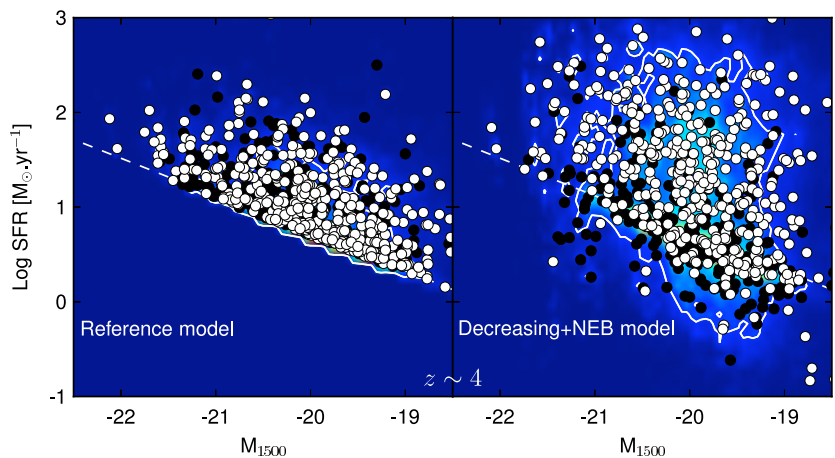}
\caption{Composite probability distribution of M$_{1500}$ and SFR for the Reference and Decreasing+NEB model for the sample at $z \sim 4$ as determined for each galaxy from our 1000 Monte Carlo simulations. The points overlaid show the median value properties for each object in the sample, black dots for Wneb and white dots for Sneb. The overlaid contour indicate the 68\% integrated probabilities on the ensemble properties measured from the centroid of the distribution. The dashed line represents the Kennicutt relation \citep{kennicutt1998}.}
  \label{debarros:fig2}
\end{figure}

Considering Wneb and Sneb, if these two populations have respectively intrinsic weak and strong nebular emission lines, we expect to find two different SFR estimations (or other difference on physical properties), with a larger SFR for Sneb. Actually, we find a difference between these two populations at each redshift. In Figure \ref{debarros:fig2}, we show for $z \sim$ 4 that Reference model doesn't permit any distinction between the two populations while with Decreasing+NEB model, Wneb objects have a median SFR lower (by a factor 4) than Sneb.  

\begin{figure}[ht!]
\centering
 \includegraphics[width=0.6\textwidth,trim=1cm 6.5cm 1cm 7cm,clip]{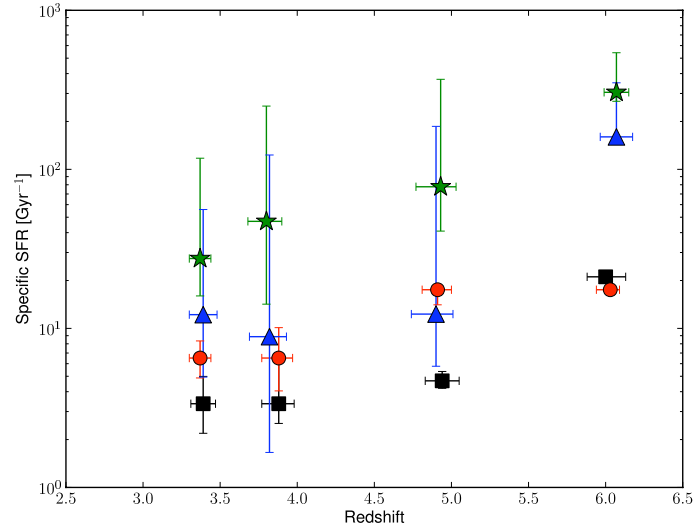}  
 \caption{Evolution of the specific SFR with redshift. Black squares: Reference model, red dots: Reference+NEB model, blue triangles: Decreasing+NEB model and green stars: Rising+NEB model. +NEB models are used for all redshift except z $\sim$ 6, for which we use +NEB+Ly$\alpha$ models.}
  \label{debarros:fig3}
\end{figure}

For example, at $z \sim$ 4, considering respectively Decreasing+NEB and Rising+NEB model, median SFR is increased by 15\% (180\%) and the median stellar mass is decreased by more than 50\% (60\%). A direct consequence is that we establish a different sSFR-redshift relation (see Figure \ref{debarros:fig3}) in comparison with previous study \citep{gonzalezetal2010}.

\section{Constraint on Ly$\alpha$}

In some models, we introduce an additionnal free parameter, a variable Ly$\alpha$ strength described by the relative Ly$\alpha$ escape fraction f$^{rel}_{Ly\alpha}$ $\epsilon$ [0,1], defined by L(Ly$\alpha$) = f$^{rel}_{Ly\alpha}$ $\times$ L$^B$(Ly$\alpha$), where L$^B$ is the intrinsic Ly$\alpha$ luminosity of the spectral template given by its Lyman continuum flux and the case B assumption and L(Ly$\alpha$) is the adopted Ly$\alpha$ luminosity for the spectral template (before any additional attenuation with the Calzetti law, assumed to affect stars and gaz in the same manner). Values f$^{rel}_{Ly\alpha}$ < 1 therefore describe an additional reduction of Ly$\alpha$ beyond the attenuation suffered by the UV continuum.

\begin{figure}[ht!]
\centering
 \includegraphics[width=0.5\textwidth,trim=0cm 5cm 1cm 3cm,clip]{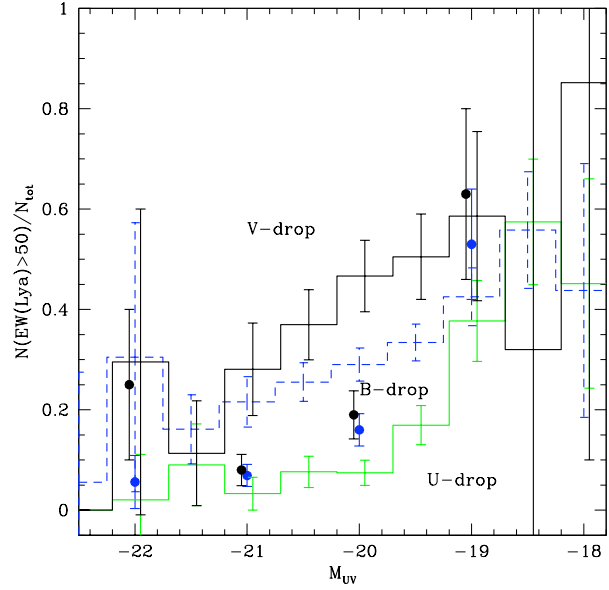}
  \caption{Fraction of galaxies with a large Ly$\alpha$ equivalent width EW(Ly$\alpha$) > 50 \AA compared to the observed fraction derived from follow-up spectroscopy of z $\sim$ 4 (blue points) and 5 (black points) LBGs (data for $B$ and $V$ drops from \cite{starketal2010})}
  \label{debarros:fig4}
\end{figure}

Overall it turns out that f$^{rel}_{Ly\alpha}$ is not well constrained for individual objects. For each sample, however, the pdf shows two relative maxima close to f$^{rel}_{Ly\alpha}$ = 0 and 1, whose relative importance varies between the samples. To quantify this behaviour further, regardless of the detailed shape of the pdf, we count the number of objects with EW(Ly$\alpha$) > 50 \AA.  We define the corresponding fraction of Ly$\alpha$ objects as r$_{Ly\alpha}$ = N(EW(Ly$\alpha$))/N$_{tot}$.  In Figure \ref{debarros:fig4}, we plot r$_{Ly\alpha}$ as a function of the absolute UV magnitude for the sample of $z \sim$ 3, 4 and 5. We find that the fraction of objects showing Ly$\alpha$ emission increases with redshift and at each redshift, we find that Ly$\alpha$ enission is more common in galaxies with fainter UV magnitudes. These two main results are in good agreement with previous spectroscopic studies \citep{starketal2011,shapleyetal2003,hayesetal2011}, showing that photometric data can also provide information on Ly$\alpha$ emission.

\section{Conclusions}

Using an updated version of the Hyperz photometric redshift code of \cite{bolzonellaetal2000} adding nebular emission (lines and continua) to the spectral templates \citep{schaerer&debarros2009,schaerer&debarros2010}, we have analysed a large sample of Lyman-break selected galaxies at $z \sim$ 3-6 in the GOODS-S field, for which deep multi-band photometry from the $U$-band to 8 $\mu$m is available.

We find that at each redshift, an exponentially decreasing star formation history with nebular emission provides statistically better SED fits than models with constant or rising star formation. Furthemore, decreasing and rising star formation with nebular emission are the only models coherent with 3.6-4.5 $\mu$m color which is likely  influenced by H$\alpha$ emission at $z \sim$ 4 \citep{shimetal2011}. These results have direct consequences on physical parameters estimation like stellar mass and SFR, which imply possibly a new mass assembly interpretation (de Barros et al. 2011, in preparation). Finally, we show that significant trends of Ly$\alpha$ strength with redshift and with UV magnitude can be inferred from broad-band photometry observations of large samples of galaxies using our models (Schaerer et al. 2011, accepted).

\begin{acknowledgements}
We acknowledge the GOODS-MUSIC collaboration. The work of SdB and DS is supported by the Swiss National Science Foundation. DPS is supported by an STFC postdoctoral research fellowship.
      \end{acknowledgements}

\bibliographystyle{aa}  
\bibliography{debarros} 

\begin{thebibliography}{18}
\expandafter\ifx\csname natexlab\endcsname\relax\def\natexlab#1{#1}\fi

\bibitem[{{Bolzonella} {et~al.}(2000){Bolzonella}, {Miralles}, \&
  {Pell{\'o}}}]{bolzonellaetal2000}
{Bolzonella}, M., {Miralles}, J., \& {Pell{\'o}}, R. 2000, \aap, 363, 476

\bibitem[{{Bruzual} \& {Charlot}(2003)}]{bruzual&charlot2003}
{Bruzual}, G. \& {Charlot}, S. 2003, \mnras, 344, 1000

\bibitem[{{Calzetti} {et~al.}(2000){Calzetti}, {Armus}, {Bohlin}, {Kinney},
  {Koornneef}, \& {Storchi-Bergmann}}]{calzettietal2000}
{Calzetti}, D., {Armus}, L., {Bohlin}, R.~C., {et~al.} 2000, \apj, 533, 682

\bibitem[{{Finlator} {et~al.}(2011){Finlator}, {Oppenheimer}, \&
  {Dav{\'e}}}]{finlatoretal2011}
{Finlator}, K., {Oppenheimer}, B.~D., \& {Dav{\'e}}, R. 2011, \mnras, 410, 1703

\bibitem[{{Gonz{\'a}lez} {et~al.}(2010){Gonz{\'a}lez}, {Labb{\'e}}, {Bouwens},
  {Illingworth}, {Franx}, {Kriek}, \& {Brammer}}]{gonzalezetal2010}
{Gonz{\'a}lez}, V., {Labb{\'e}}, I., {Bouwens}, R.~J., {et~al.} 2010, \apj,
  713, 115

\bibitem[{{Hayes} {et~al.}(2011){Hayes}, {Schaerer}, {{\"O}stlin}, {Mas-Hesse},
  {Atek}, \& {Kunth}}]{hayesetal2011}
{Hayes}, M., {Schaerer}, D., {{\"O}stlin}, G., {et~al.} 2011, \apj, 730, 8

\bibitem[{{Kennicutt}(1998)}]{kennicutt1998}
{Kennicutt}, Jr., R.~C. 1998, \araa, 36, 189

\bibitem[{{Madau}(1995)}]{madau1995}
{Madau}, P. 1995, \apj, 441, 18

\bibitem[{{Ono} {et~al.}(2010){Ono}, {Ouchi}, {Shimasaku}, {Dunlop}, {Farrah},
  {McLure}, \& {Okamura}}]{onoetal2010}
{Ono}, Y., {Ouchi}, M., {Shimasaku}, K., {et~al.} 2010, \apj, 724, 1524

\bibitem[{{Salpeter}(1955)}]{salpeter1955}
{Salpeter}, E.~E. 1955, \apj, 121, 161

\bibitem[{{Santini} {et~al.}(2009){Santini}, {Fontana}, {Grazian}, {Salimbeni},
  {Fiore}, {Fontanot}, {Boutsia}, {Castellano}, {Cristiani}, {de Santis},
  {Gallozzi}, {Giallongo}, {Menci}, {Nonino}, {Paris}, {Pentericci}, \&
  {Vanzella}}]{santinietal2009}
{Santini}, P., {Fontana}, A., {Grazian}, A., {et~al.} 2009, \aap, 504, 751

\bibitem[{{Schaerer} \& {de Barros}(2009)}]{schaerer&debarros2009}
{Schaerer}, D. \& {de Barros}, S. 2009, \aap, 502, 423

\bibitem[{{Schaerer} \& {de Barros}(2010)}]{schaerer&debarros2010}
{Schaerer}, D. \& {de Barros}, S. 2010, \aap, 515, A73+

\bibitem[{{Shapley} {et~al.}(2003){Shapley}, {Steidel}, {Pettini}, \&
  {Adelberger}}]{shapleyetal2003}
{Shapley}, A.~E., {Steidel}, C.~C., {Pettini}, M., \& {Adelberger}, K.~L. 2003,
  \apj, 588, 65

\bibitem[{{Shim} {et~al.}(2011){Shim}, {Chary}, {Dickinson}, {Lin}, {Spinrad},
  {Stern}, \& {Yan}}]{shimetal2011}
{Shim}, H., {Chary}, R.-R., {Dickinson}, M., {et~al.} 2011, \apj, 738, 69

\bibitem[{{Stark} {et~al.}(2009){Stark}, {Ellis}, {Bunker}, {Bundy}, {Targett},
  {Benson}, \& {Lacy}}]{starketal2009}
{Stark}, D.~P., {Ellis}, R.~S., {Bunker}, A., {et~al.} 2009, \apj, 697, 1493

\bibitem[{{Stark} {et~al.}(2010){Stark}, {Ellis}, {Chiu}, {Ouchi}, \&
  {Bunker}}]{starketal2010}
{Stark}, D.~P., {Ellis}, R.~S., {Chiu}, K., {Ouchi}, M., \& {Bunker}, A. 2010,
  \mnras, 408, 1628

\bibitem[{{Stark} {et~al.}(2011){Stark}, {Ellis}, \& {Ouchi}}]{starketal2011}
{Stark}, D.~P., {Ellis}, R.~S., \& {Ouchi}, M. 2011, \apjl, 728, L2+

\end{thebibliography}

\end{document}